\documentclass[cmbright]{WileySTAT-V1} %% [cmbright,doublespace], [demo] for draft mode

\usepackage[bookmarksopen,bookmarksnumbered,citecolor=blue,urlcolor=blue]{hyperref} %% Online requirements
\usepackage{natbib}     %% Used for Author Name and year

\usepackage{dcolumn}    %% For Tabular decimal alignments
\usepackage{dsfont}
\usepackage[utf8]{inputenc}
\usepackage{amsthm,amsmath}
\usepackage{bm}
\usepackage{bbm}
\usepackage{moreverb}
\usepackage{booktabs}
\usepackage{tikz}
\usepackage{subfig}
\usepackage{pgfplots}
\usepackage{xcolor}
\usepackage{logicpuzzle}
\usepackage{tablefootnote}
\usepackage{amstext}
\usepackage{amsfonts}

\newcolumntype{d}[1]{D{.}{.}{#1}}

\DeclareMathOperator{\logit}{logit}

\def\volumeyear{}
\def\volumenumber{}
\def\DOI{}

\runninghead{Remiro-Az\'ocar \textsc{et al}}{
Marginal and conditional summary measures: transportability and compatibility}

\theoremstyle{plain}
\begin{document}

\title{Marginal and conditional summary measures: transportability and compatibility across studies}

\author{Antonio Remiro-Az\'ocar\affil{a}\corrauth, David M. Phillippo\affil{b}, Nicky J. Welton\affil{b}, Sofia Dias\affil{c}, A. E. Ades\affil{b}, Anna Heath\affil{d,e,f} and Gianluca Baio\affil{f}}

\address{%
\affilnum{a}Methods and Outreach, Novo Nordisk Pharma, Madrid, Spain \\
\affilnum{b}Bristol Medical School (Population Health Sciences), University of Bristol, Bristol, United Kingdom\\
\affilnum{c}Centre for Reviews and Dissemination, University of York, York, United Kingdom\\
\affilnum{d}Child Health Evaluative Sciences, The Hospital for Sick Children, Toronto, Canada \\
\affilnum{e}Dalla Lana School of Public Health, University of Toronto, Toronto, Canada\\
\affilnum{f}Department of Statistical Science, University College London, London, United Kingdom}

\articlenote{}
\corremail{aazw@novonordisk.com}

\begin{abstract}
Marginal and conditional summary measures do not generally coincide, have different interpretations and correspond to different decision questions. While these aspects have primarily been recognized for non-collapsible summary measures, they are equally problematic for some collapsible measures in the presence of effect modification. We clarify the interpretation and properties of several marginal and conditional summary measures, considering different types of outcomes and hypothetical outcome-generating mechanisms. We describe implications of the choice of summary measure for transportability, highlighting that covariates not conventionally described as effect modifiers can modify population-level treatment effects. Finally, we illustrate existing summary measure incompatibility issues in the context of evidence synthesis, using the case of covariate adjustment methods for indirect treatment comparisons. Because marginal and conditional summary measures do not generally coincide, their na\"ive pooling in evidence synthesis can produce bias. Almost invariably, care is needed to ensure that evidence synthesis methods are combining compatible summary measures, and this may be easier to accomplish with full access to individual patient data.
\end{abstract}


\maketitle

\section{Introduction}\label{sec1}

Randomized controlled trials (RCTs) are the ``gold standard'' research design for estimating causal treatment effects with high internal validity. Randomization ensures that treatment assignment is independent of all measured and unmeasured baseline covariates, thereby guaranteeing that these are balanced in expectation over the trial's treatment arms. Nevertheless, high internal validity does not necessarily guarantee external validity with respect to a target population of substantive interest. External validity is threatened by differences in the distribution of effect modifiers; that is, covariates inducing differential treatment effects between the RCT and the target population \citep{westreich2019target}. 

Differences in the distribution of effect modifiers can compromise the \textit{transportability} or \textit{constancy} of treatment effects across populations \citep{greenland2014effect, mcknight2015effect}. While such differences are recognized as a fundamental component of treatment effect heterogeneity, a lesser-known source of heterogeneity is the summary (effect) measure used to compare outcomes between treatment arms. Summary measures can be marginal (unconditional) or conditional; directly collapsible (mean and risk difference), collapsible but not directly collapsible (risk ratio) or non-collapsible (odds ratio and hazard ratio) \citep{greenland2015collapsibility}. Definitions of these terms are provided in Table \ref{key_definitions}. 

Marginal and conditional summary measures do not generally coincide and correspond to different decision questions, with discussion over which is more suitable for healthcare decision-making \citep{phillippo2021target, remiro2022target, Phillippo_Remiro-Azócar_Heath_Baio_Dias_Ades_Welton_2025}. Whether a summary measure is marginal or conditional has implications on its transportability across populations. In evidence synthesis, the na\"ive pooling of marginal and conditional measures can produce bias \citep{remiro2021methods, remiro2022parametric}, particularly for analyses using aggregate-level data, as trials may report different summary measures. This issue of ``estimand incompatibility'' has primarily been recognized for non-collapsible summary measures, where marginal and conditional measures do not coincide, even in the absence of effect modification by the covariates \citep{daniel2021making}. The issue is equally problematic for some collapsible measures – those that are not directly collapsible – in the presence of effect modification \citep{remiro2024transportability}.

In this article, we clarify the interpretation of different marginal and conditional summary measures. We do not intend to express a preference towards one or another, but to set out the relevant terminology and illustrate the properties of relevant summary measures. We consider different types of outcomes and outcome-generating mechanisms, with and without effect modification. We describe implications of the choice of summary measure for transportability, highlighting that covariates not conventionally described as effect modifiers can modify population-level treatment effects. We conclude by illustrating existing summary measure incompatibility issues in the context of evidence synthesis, using the case of covariate adjustment methods for indirect treatment comparisons. Care has to be exercised to ensure that evidence synthesis methods combine compatible summary measures. 

\newcommand{\specialcell}[2][c]{%
  \begin{tabular}[#1]{@{}l@{}}#2\end{tabular}}

\begin{table}[!htb]
\caption{Definitions of key terminology for summary measures.}
\centering
\scalebox{0.83}{
\begin{tabular}{ll}
\toprule
Term & Definition \\
\midrule
Collapsible & \specialcell[t]{Summary measure for which the marginal measure can always be expressed as a weighted average of individual- \\ or subgroup-level conditional measures} \\
Conditional & \specialcell[t]{Summary measure contrasting (functions of) the conditional outcome distributions under different treatments, \\ given specific covariate values} \\
Directly collapsible & \specialcell[t]{Collapsible summary measure for which the collapsibility weights are given by the marginal distributions of the \\ covariates that are conditioned on, e.g., proportions for discrete covariates \citep{colnet2023risk}} \\
Marginal & \specialcell[t]{Summary measure contrasting (functions of) the marginal or unconditional outcome distributions under \\ different treatments} \\
Non-collapsible & \specialcell[t]{Summary measure for which the marginal measure does not equal a weighted average of conditional measures, \\ even where the conditional measure is equal across all individuals or subgroups \citep{huitfeldt2019collapsibility}} \\
\bottomrule
\end{tabular}
}
\label{key_definitions}
\end{table}

\section{Marginal and conditional estimands}\label{sec2}

An estimand, as defined by the International Council of Harmonisation (ICH) E9 (R1) Addendum, is a precise definition of the treatment effect that is targeted by a clinical trial \citep{keene2023estimands}. The notion of an estimand as a target of estimation is not new, but its application to pose research questions in registrational RCTs has been encouraged by ICH E9 (R1). The summary measure is only one of the five attributes of estimands, as per ICH E9 (R1). Nevertheless, because our focus is on such attribute, we will use the terms summary measure and estimand interchangeably. Following the potential outcomes framework, these are defined as causal contrasts of any function of the distributions of counterfactual outcomes under different treatments for the same set of subjects \citep{hernan2020causal}. Because ICH E9 (R1) is aligned with counterfactual language, potential outcomes notation is useful to formally construct the summary measures.

% \footnote{Another central component of estimands in ICH E9 (R1) is the intercurrent event strategy. We shall assume that a treatment policy strategy is followed, closely related to the intention-to-treat principle, such that the occurrence of intercurrent events, e.g.,~treatment discontinuation or switching, is deemed irrelevant to define the estimands of interest.}

Consider a hypothetical ideal RCT targeting a population of substantive interest. Let $Y^t$ denote the potential outcome that might have been observed for an individual assigned to binary treatment $T=t$, with $t \in \{0, 1\}$ indicating assignment to the control and the active treatment group, respectively. Each participant in the trial has two potential outcomes, $Y^1$ and $Y^0$. Assuming no loss to follow up and complete data, one of these is observed in the trial. The other is the outcome hypothetically realized under the other treatment.

The \textit{marginal treatment effect} (MTE) estimand is a contrast of functions of marginal or unconditional outcome expectations under each treatment. On the additive scale:
\begin{equation}
\textnormal{MTE} = g\left(E\left(Y^1 \right )\right) - g\left(E\left(Y^0 \right)\right),
\label{additive}
\end{equation}
where $g(\cdot)$ is a suitable ``link'' function, transforming the potential outcome means onto the ``minus/plus infinity range'', and each expectation $E(\cdot)$ is over the distribution of potential outcomes in the trial population. The marginal estimand is typically described as the average treatment effect in the trial population, had everyone in the trial population been assigned active treatment versus control \citep{austin2014use}. Note that the use of a link function is not strictly necessary for the definition and is usually obviated in the causal inference literature. 

The marginal estimand is often considered of primary interest for population-level policy decisions \citep{remiro2022target}. As per Equation \ref{additive}, the causal inference literature tends to define population effects through functions of the marginal distributions of outcomes, although ``the word marginal is a preference rather than necessary to the definition'' because marginal estimands appear to be favored for causal inference based on observational data \citep{morris2022planning}. Consequently, the marginal treatment effect is commonly referred to as the \textit{average treatment effect} (ATE) or the \textit{population-average treatment effect} (PATE) \citep{ austin2013performance, remiro2024transportability}, but we will not use this terminology to avoid confusion with other (population-)average treatment effects introduced later in this section. 

\textit{Conditional treatment effect} estimands contrast functions of outcome expectations that are conditional on baseline covariates, with these playing an explicit role in the treatment effect definition. There are different conditional estimands that are potentially of interest to researchers, with different interpretations and corresponding to different decision problems. We make a conceptual distinction between three conditional summary measures: the \textit{conditional treatment effect} evaluated at particular covariate values, the \textit{conditional treatment effect at the (covariate) means}, and the \textit{population-average conditional treatment effect}. 

While, by virtue of randomization, the marginal estimand within the ideal RCT can be identified from the observed data with minimal assumptions, the identification of conditional estimands within such RCT may require statistical assumptions about model validity, particularly with continuous covariates or where there are only a few subjects in some covariate strata \citep{van2022estimands, van2023use}.

We define the conditional treatment effect (CTEX) given a particular covariate value $X=x$ as: 
\begin{equation}
\textnormal{CTEX}
= g\left(E\left(Y^1 \mid X=x \right)\right) - g\left(E\left(Y^0 \mid X=x \right)\right),
\label{cond_additive}
\end{equation}
where each expectation $E(\cdot)$ is over the distribution of potential outcomes in the trial population, but conditional on covariate values. The CTEX is the average treatment effect had a subset of individuals, with covariate profile $X=x$, been assigned active treatment versus control. It is often referred to as the \textit{conditional average treatment effect} (CATE) \citep{colnet2023risk}, but we will not use this term to avoid confusion with other average conditional effects introduced later in this section.

To simplify the exposition, we assume that there is a single baseline covariate. If this is binary or categorical, the CTEX can be interpreted as a subgroup- or stratum-specific effect. If multiple relevant covariates are conditioned on, it comes closer to an individualized or subject-specific effect. The CTEX is of interest to clinicians and individual patients, particularly if subjects with different covariate values benefit differently from treatment. It is arguably of lesser interest for population-level decision-making, especially if the treatment effect does not change with the covariate values, but can be useful when considering decisions for different subgroups or subpopulations if the treatment effect does vary.   

The conditional treatment effect at the (covariate) means (CTEM) is a special case of the CTEX:
\begin{equation}
\textnormal
{CTEM} =  g\left(E\left(Y^1 \mid X=\bar{X} \right)\right) - g\left(E\left(Y^0 \mid X=\bar{X} \right)\right),  
\label{ctem}
\end{equation}
with $\bar{X}=E(X)$. This reflects the conditional treatment effect for subjects with covariate values equal to their mean. While choice of the mean is somewhat arbitrary, as opposed to the median or any other statistic, this summary measure may be of interest if the covariate is continuous. Nevertheless, it is sensitive to the scale of the covariates, not generally a desirable target for population-level decision-making and likely not meaningful for discrete covariates. 

Consider sex, ethnicity, smoking status, biomarker status or exposure to previous treatment, where the mean is the proportion of subjects in a category. The CTEM does not make much sense and may have an awkward or confusing interpretation, denoting the effect for a non-existent ``average subject'' in the middle of the categories. Confusingly, the CTEM and conditional outcome expectations $E\left(Y^t \mid X=\bar{X} \right)$ for $t \in \{ 0, 1\}$ are sometimes known as marginal effects (at the mean) and marginal (or ``least-squares'') means, respectively, in the econometrics literature \citep{arel2024interpret, lenth2016least}.
  
Finally, the population-average conditional treatment effect (PACTE) is defined as:
\begin{align}
\textnormal{PACTE}
&= E_X \left (
\textnormal{CTEX}
\right ) \nonumber \\
&= E_X 
\left (
g\left(E\left(Y^1 \mid X=x \right)\right) - g\left(E\left(Y^0 \mid X=x \right)\right)  
\right),
\label{pacte}
\end{align}
where each inner expectation $E(\cdot)$ is over the distribution of potential outcomes in the trial population, conditional on covariate values, and the outer expectation $E_X(\cdot)$ is over the trial population. Namely, the PACTE is the average CTEX across all the subjects or subgroups in the trial population. While the terms population-average and marginal are often used interchangeably, the PACTE is also a population-average summary measure because it applies to the whole trial population and is an average over the covariate distribution of such population. 

It has been argued that the PACTE can be interpreted as the conditional treatment effect at the (covariate) means \citep{remiro2021conflating}. Nevertheless, as we shall discuss in Section \ref{sec3}, the PACTE coincides with the CTEM only when the CTEX is a linear function of $X$ (or constant). The PACTE and the CTEM are fundamentally distinct summary measures. Note that while the CTEX is a function of the covariate(s), the CTEM and the PACTE (and the MTE) compress the treatment effect surface into a single aggregate summary. 

The difference between the MTE and the PACTE is the order of operations, or the scale on which the averaging takes place \citep{Phillippo_Remiro-Azócar_Heath_Baio_Dias_Ades_Welton_2025}. The MTE takes the unconditional expectation of the potential outcomes on their natural scale, then contrasts the transformed averages on the scale imposed by the link function: 
\begin{align*}
\textnormal{MTE} &= g\left(E\left(Y^1 \right )\right) - g\left(E\left(Y^0 \right)\right) \\ 
&=
g
\left ( E_X\left(E\left(Y^1 \mid X=x
\right )\right) \right )
- g \left (E_X   \left(E\left(Y^0 \mid X=x
\right )\right) \right ), 
\end{align*}
by the law of total expectation. Conversely, the PACTE contrasts the conditional expectations of the potential outcomes on the scale imposed by the link function, then takes the average of the contrasts, with the outer expectation over $X$ outside the link functions. The changing order of operations is irrelevant when $g(\cdot)$ is the identity link, but otherwise matters.

\section{Model-based estimands}\label{sec3}

The summary measure definitions in Equations \ref{additive} to \ref{pacte} are formulated on an additive scale. Such definitions are ``model-free''; despite the use of function $g(\cdot)$, the estimands do not necessarily rely on modeling assumptions and are not necessarily encoded by the coefficients of a model \citep{van2023use, vansteelandt2022assumption, mutze2025principles}. Nevertheless, in practice, the choices of $g(\cdot)$ and the scale of the summary measure are often ``model-based'': influenced by modeling preferences or statistical considerations for different outcome types \citep{van2023use, daniel2021making, mutze2025principles}. 

For instance, the logit link function is used to model binary outcomes because fitted logistic regressions predict outcome probabilities between zero and one. For non-negative integer (``count'') outcomes, the logarithmic link naturally incorporates the lower bound of a count at zero. Typically, within a generalized linear modeling (GLM) framework: for continuous, count or binary outcomes, the link function $g(\cdot)$ is a \textit{canonical} link; the identity, logarithmic or logit link, respectively, with the mean difference, log rate ratio or log odds ratio scale, respectively, as the additive linear predictor scale imposed by the link function of the GLM \citep{wei2024conditional, hojbjerre2025powering}. 

% Nevertheless, this is rarely carried out in practice in evidence synthesis, where a model-based view of estimands is prevalent \citep{van2013evidence, caldwell2012selecting}. 

Admittedly, the choice of scale for the comparison is up to the analyst, informed by models as well as value judgments from substantive experts. One can use a logit link function to model binary outcomes with a logistic regression, then report the treatment effect as a difference of (marginal) risks, using an identity link for summarization. Nevertheless, as is often the case in practice for RCTs \citep{wei2024conditional, mutze2025principles}, we shall assume that the scale for the target summary measure is determined by the GLM canonical link function that would be used to model the outcome type. Additionally, we suppose that the conditional outcome expectations follow particular parameterizations with specific functional forms about the effect of covariates on conditional measures. This implies that the conditional estimands are known and can be expressed in simple terms composed of model parameter(s). In reality, the ``true'' generative function for the conditional outcome expectation will be more complex, and one is likely to have incomplete knowledge on functional forms for the baseline covariates. 

Under the GLM framework, we postulate three hypothetical parametric outcome-generating models to explore the behavior of estimands under different generative mechanisms. These are not exhaustive and are posited for illustrative purposes. The first outcome-generating mechanism induces treatment effect homogeneity across subjects, such that the CTEX on the linear predictor scale is constant. In this scenario, the MTE, CTEM and PACTE are equivalent for collapsible summary measures, but the MTE does not coincide with any of the conditional quantities for non-collapsible summary measures. 

The second outcome-generating mechanism induces treatment effect heterogeneity with the CTEX varying as a linear function of $X$. In this scenario, the MTE, CTEM and PACTE are equivalent for directly collapsible summary measures, but the MTE does not coincide with any of the conditional quantities for summary measures that are not directly collapsible. The third outcome-generating mechanism also induces treatment effect heterogeneity but with the CTEX varying as a non-linear -- assumed quadratic -- function of $X$. In this case, the CTEM and the PACTE are not equivalent, regardless of whether there is collapsibility (or direct collapsibility) or not.

\subsection{Homogeneous CTEX}\label{subsec31}

The outcome model with treatment effect homogeneity, called the \textit{homogeneous} illustrative model, specifies the conditional expectation of potential outcome $Y^t$ for treatment $T=t$ given $X$ as:
\begin{equation}
E (Y^{t} \mid X) = g^{-1} \left( \beta_0 + \beta_X X + 
\beta_T t \right),  
\label{eqn1}
\end{equation}
where $g(\cdot)$ is a suitable link function, $\beta_0 \in \mathbb{R}$ is an intercept quantifying the ``baseline risk'' on the linear predictor scale (the conditional outcome expectation on such scale for $X=0$ and $T=0$), and the model coefficients $\beta_X, \beta_T \in \mathbb{R}$ quantify conditional predictor-outcome associations. We place no further statistical assumptions, e.g.,~normality, on the conditional outcome distribution. Because the outcome model does not include a covariate-treatment product (``interaction'') term but does include a term for the main covariate effect, covariate $X$ is said to be purely \textit{prognostic} \citep{ballman2015biomarker}. The baseline risk $\beta_0$ is often a proxy for unobserved prognostic factors (assuming that $X=0$ and $T=0$ correspond to a suitable reference value).

For any link function (the link plays no role in the calculus), substituting Equation \ref{eqn1} into Equation \ref{cond_additive} yields $\textnormal{CTEX} = \beta_T$ on the linear predictor scale. In this case, the CTEX is not actually covariate-specific; the same CTEX applies to every individual and conceivable subgroup, and we have treatment effect \textit{homogeneity}. The conditional treatment effect evaluated at the (covariate) mean $\textnormal{CTEM}=\beta_T$ is the same as for every other value of the covariate. Because the PACTE averages the constant CTEX across the subjects in the population, $\textnormal{PACTE}=\beta_T$.  

With respect to the marginal estimand, suppose $g(\cdot)$ is the identity link, such that the outcome-generating model is linear. On the mean difference scale, by the law of total expectation, the marginal treatment effect estimand $\textnormal{MTE} = \beta_T$, and the model-based coefficient $\beta_T$ coincides with the model-free definition of the marginal estimand and all conditional estimands \citep{remiro2024transportability}. With the homogeneous linear generative model, the marginal mean difference does not depend on the distribution of the purely prognostic covariate $X$. 

Now, suppose $g(\cdot) = \log(\cdot)$, such that the outcome-generating mechanism is log-linear; for instance, a Poisson or negative binomial model. These models also add the logarithm of person-time, $\log(\tau)$, as an offset to the right-hand side of Equation \ref{eqn1}. We shall assume that person-time is fixed across subjects, e.g., corresponding to the trial follow-up period, but we return to this caveat later in this section. On the log rate ratio scale, $\textnormal{MTE} = \beta_T$, and the model-based coefficient $\beta_T$ coincides with the model-free definition of the marginal estimand and all conditional estimands \citep{remiro2024transportability, vellaisamy2008collapsibility}. With the homogeneous log-linear generative model (and constant person-time), the marginal (log) rate ratio does not depend on the distribution of the purely prognostic covariate $X$. 

Finally, suppose $g(\cdot) = \logit(\cdot)$, such that the outcome-generating model is logistic. Because the (log) odds ratio is \textit{non-collapsible} \citep{daniel2021making}, the marginal summary measure cannot be expressed as a weighted average of individual- or subgroup-level CTEX, not even in the absence of treatment effect heterogeneity \citep{colnet2023risk, huitfeldt2019collapsibility}. When $\beta_X \neq 0$, the marginal and conditional (log) odds ratios are, almost invariably, not equal, with the marginal estimand closer to the null hypothesis -- zero on the log odds ratio scale and one on the odds ratio scale -- than all the conditional estimands. 

Via Jensen's inequality and the monotonicity of the logit function, it can be shown that: (1) $\beta_T > \textnormal{MTE}$ if $\beta_T > 0$; and (2) $\beta_T < \textnormal{MTE}$ if $\beta_T < 0$ \citep{colnet2023risk}. While the model-based coefficient $\beta_T$ can be interpreted as a conditional estimand, it cannot be interpreted as a population-average marginal estimand because it is not equal to the model-free definition of the $\textnormal{MTE}$, despite enforcing treatment effect homogeneity across subjects on the (log) odds ratio scale. Moreover, the $\textnormal{MTE}$ on such scale does not only depend on $\beta_T$ nor on the mean of purely prognostic covariate $X$. It depends on the full distribution of $X$, as well as the distribution of baseline risk because prognostic factors may not even be known or measured \citep{Phillippo_Remiro-Azócar_Heath_Baio_Dias_Ades_Welton_2025}. 

Mean differences, risk differences and risk ratios are collapsible. One would think that (log) rate ratios inherit collapsibility from (log) risk ratios, as rates are risks per unit time. However, (log) rate ratios are non-collapsible \citep{sjolander2016note}, unless person-time is not influenced by the outcomes (treatment or baseline covariates), for instance, in a study of non-fatal recurrent events where all subjects remain at risk throughout the trial follow-up period. We will assume that this is the case for ease of exposition; this allows us to describe (log) rate ratios as (log) risk ratios when discussing count outcomes and log-linear outcome models.  

\begin{figure}[!htb]
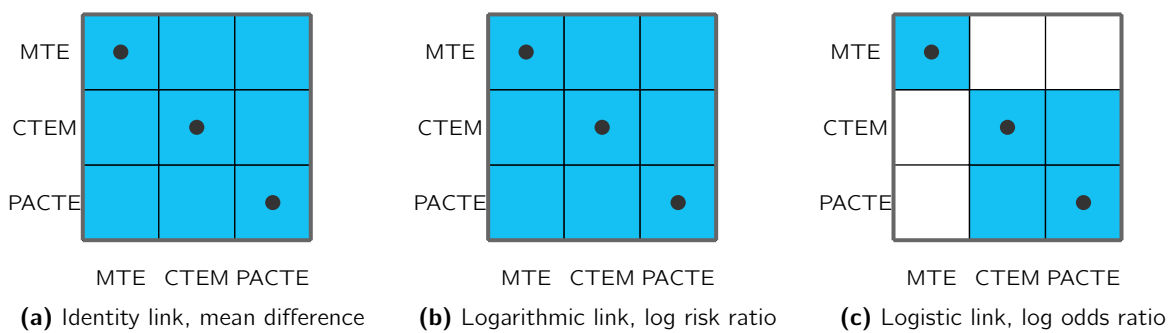

\centering
\subfloat[Identity link, mean difference]{
\begin{logicpuzzle}[rows=3,columns=3,color=cyan!70, fontsize=small]
\valueH{MTE, CTEM, PACTE}
\valueV{PACTE,CTEM,MTE}
\fillcell{1}{1}
\fillcell{1}{2}
\fillcell{1}{3}
\fillcell{2}{1}
\fillcell{2}{2}
\fillcell{2}{3}
\fillcell{3}{1}
\fillcell{3}{2}
\fillcell{3}{3}
\setrow{3}{{\Graveltrap},{},{}}
\setrow{2}{{},{\Graveltrap},{}}
\setrow{1}{{},{},{\Graveltrap}}
\framepuzzle[black!60]
\end{logicpuzzle}
}
\subfloat[Logarithmic link, log risk ratio]{
\begin{logicpuzzle}[rows=3,columns=3,color=cyan!70, fontsize=small]
\valueH{MTE, CTEM, PACTE}
\valueV{PACTE,CTEM,MTE}
\fillcell{1}{1}
\fillcell{1}{2}
\fillcell{1}{3}
\fillcell{2}{1}
\fillcell{2}{2}
\fillcell{2}{3}
\fillcell{3}{1}
\fillcell{3}{2}
\fillcell{3}{3}
\setrow{3}{{\Graveltrap},{},{}}
\setrow{2}{{},{\Graveltrap},{}}
\setrow{1}{{},{},{\Graveltrap}}
\framepuzzle[black!60]
\end{logicpuzzle}
}
\subfloat[Logistic link, log odds ratio]{
\begin{logicpuzzle}[rows=3,columns=3,color=cyan!70, fontsize=small]
\valueH{MTE, CTEM, PACTE}
\valueV{PACTE,CTEM,MTE}
\fillcell{1}{3}
\fillcell{2}{1}
\fillcell{2}{2}
\fillcell{3}{1}
\fillcell{3}{2}
\setrow{3}{{\Graveltrap},{},{}}
\setrow{2}{{},{\Graveltrap},{}}
\setrow{1}{{},{},{\Graveltrap}}
\framepuzzle[black!60]
\end{logicpuzzle}
}
\caption{Matrices indicating whether different estimands are equivalent for the homogeneous illustrative models. The blue squares denote matching estimand values; the dots denote the diagonal, where estimands are equivalent by definition.}
\label{matrices_homogeneous}
\end{figure}

Table \ref{estimands_homogeneous} presents a taxonomy of marginal estimands based on the homogeneous illustrative models. Figure \ref{matrices_homogeneous} indicates whether different estimands match for the homogeneous illustrative models, across link functions and summary measures. The CTEX is not displayed because, outside the homogeneous illustrative model, it is a function of the covariates and not a ``one number'' summary. The blue squares denote matching estimand values; the dots denote the diagonal, where logically the estimands are equivalent by definition. In Table \ref{estimands_homogeneous} and Figure \ref{matrices_homogeneous}, the summary measures are on the additive linear predictor scale imposed by the link function used to model the corresponding outcome type. 

\begin{table}[!htb]
\caption{Model-based marginal estimands for the homogeneous illustrative models. For count outcomes and the log link, person-time is assumed constant, such that the log rate ratio is collapsible and can be interpreted a log risk ratio.}
\scalebox{0.9}{
\begin{tabular}{llll}
\toprule
Outcome & Link function & Summary measure & 
Marginal estimand \\
\midrule
Continuous & Identity & Mean difference & Does not depend on the distribution of purely prognostic covariates \\
Count & Logarithmic & Log risk ratio & Does not depend on the distribution of purely prognostic covariates \\
Binary & Logit & Log odds ratio & Depends on the full joint distribution of purely prognostic covariates \\
\bottomrule
\end{tabular}
}
\label{estimands_homogeneous}
\end{table}

Outside the GLM framework, a model and summary measure that are widely used for the analysis of time-to-event outcomes in RCTs are the Cox proportional hazards model and the (log) hazard ratio. While the Cox model uses a logarithmic link to map hazards to the linear predictor, (log) hazard ratios are non-collapsible owing to conditioning on past survival, similar to (log) rate ratios \citep{sjolander2016note}. The marginal (log) hazard ratio depends on the shape of the baseline hazard and on the full joint distribution of purely prognostic covariates that do not ``interact'' with treatment, even under a generative model enforcing the constancy of the conditional (log) hazard ratio across covariate values \citep{martinussen2013collapsibility, daniel2021making, Phillippo_Remiro-Azócar_Heath_Baio_Dias_Ades_Welton_2025}. Such generative model would result in non-proportional hazards for the marginal summary measure because proportional hazards cannot hold simultaneously on the marginal and conditional scales \citep{daniel2021making, Phillippo_Remiro-Azócar_Heath_Baio_Dias_Ades_Welton_2025}. 

In summary, while the \textit{within-trial} estimation of the MTE generally requires weaker statistical assumptions than the estimation of the conditional summary measures, it likely requires stronger assumptions for transportability across different populations. This is particularly the case for non-collapsible summary measures, given their MTE dependence on the joint distribution of purely prognostic covariates, even in the absence of treatment effect heterogeneity.

\subsection{Linear heterogeneous CTEX}\label{subsec32}

The homogeneous illustrative model in Section \ref{subsec31} is sometimes too restrictive, in which case one can postulate an outcome model that accommodates for treatment effect heterogeneity across individuals. The \textit{heterogeneous} illustrative model specifies the conditional expectation of potential outcome $Y^t$ for treatment $T=t$ given $X$ as:
\begin{equation}
E (Y^{t} \mid X) = g^{-1} \left( \beta_0 + \beta_X X + \beta_T t + \beta_{XT}X t \right). 
\label{eqn1_interaction}
\end{equation}
Here, $X$ is prognostic of outcome, with the coefficient $\beta_X \in \mathbb{R}$ quantifying the conditional covariate-outcome association in the control group. 

Assuming that the interaction coefficient $\beta_{XT} \in \mathbb{R}$ is non-null, $X$ also modifies the (conditional) treatment effect on the linear predictor scale. In this case, covariate $X$ is said to be an \textit{effect modifier}. The RCT literature has typically conceptualized effect modifiers in this way, as baseline covariates that induce treatment effect heterogeneity through treatment-covariate interactions in an outcome-generating model \citep{christensen2021effect}. 

Substituting Equation \ref{eqn1_interaction} into Equation \ref{cond_additive} gives $\textnormal{CTEX} = \beta_T + \beta_{XT} x$ at $X=x$ on the linear predictor scale. That is, for the identity, logarithmic or logit link, the corresponding conditional mean difference, log risk ratio or log odds ratio, respectively, depends on the level of $X$. As such, $\beta_T$ is only the conditional treatment effect evaluated at $X=0$, and there is no longer a single CTEX estimand. Substituting Equation \ref{eqn1_interaction} into Equation \ref{ctem} yields $\textnormal{CTEM}=\beta_T + \beta_{XT} \bar{X}$. Similarly, substituting Equation \ref{eqn1_interaction} into Equation \ref{pacte} gives $\textnormal{PACTE}=E_X \left (\beta_T + \beta_{XT} X \right ) = \beta_T + \beta_{XT} E_X \left (  X  \right ) = \beta_T + \beta_{XT}\bar{X}$.  

With respect to the marginal estimand, suppose $g(\cdot)$ is the identity link, such that the outcome-generating model is linear. On the mean difference scale, by the law of total expectation, $\textnormal{MTE} = \beta_T + \beta_{XT}\bar{X}$. While $\beta_T$ no longer has a marginal interpretation, the marginal estimand coincides with the CTEM and PACTE, and can still be expressed in terms of model coefficients and the unconditional expectation of $X$. This is because the mean difference is directly collapsible \citep{huitfeldt2019collapsibility}. Namely, the marginal measure is equal to a weighted average of conditional measures, with the weights given by marginal distributions of the covariates that are conditioned on, e.g., for binary or categorical covariates, the weights would be the covariate proportions \citep{colnet2023risk}. 

Unlike the mean difference, the (log) risk ratio is not directly collapsible \citep{kiefer2019average}. The collapsibility weights for this summary measure are more complex; the marginal (log) risk ratio is in general not equal to a weighted average of conditional (log) risk ratios, if the weights are given by marginal covariate summary moments \citep{huitfeldt2019collapsibility}. Consequently, with $g(\cdot) = \log(\cdot)$ and the heterogeneous log-linear generative model, the marginal (log) risk ratio cannot be expressed in terms of $\beta_T$, $\beta_{XT}$ and $\bar{X}$. In this scenario, the MTE on the log risk ratio scale is not equivalent to the CTEM or the PACTE \citep{remiro2024transportability}.  

Moreover, when there are multiple prognostic baseline covariates and in the presence of treatment effect heterogeneity, the marginal (log) risk ratio cannot be identified using the marginal distributions of the effect modifiers \citep{remiro2024transportability}. Rather, one requires the full joint (multivariate) distribution of the effect modifiers and the purely prognostic covariates that are associated with the former. For non-collapsible measures such as the (log) odds ratio, the MTE depends on the full joint distribution of purely prognostic covariates, even if these are not associated with any effect modifiers or in the absence of treatment effect heterogeneity. 

\begin{figure}[!htb]
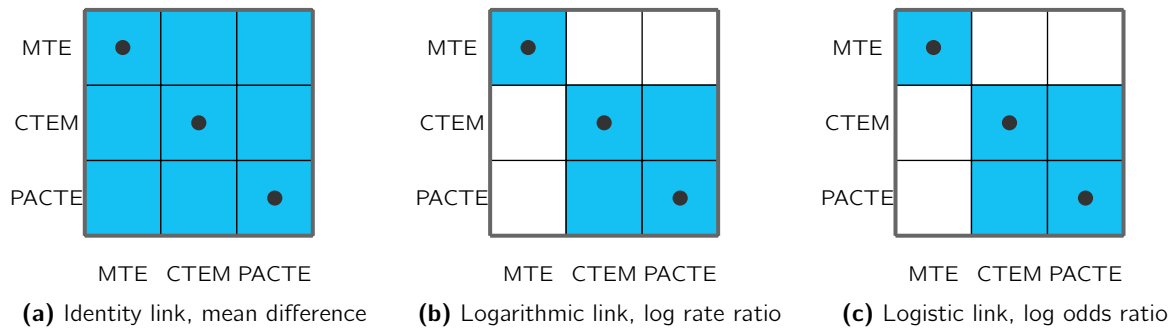

\centering
\subfloat[Identity link, mean difference]{
\begin{logicpuzzle}[rows=3,columns=3,color=cyan!70, fontsize=small]
\valueH{MTE, CTEM, PACTE}
\valueV{PACTE,CTEM,MTE}
\fillcell{1}{1}
\fillcell{1}{2}
\fillcell{1}{3}
\fillcell{2}{1}
\fillcell{2}{2}
\fillcell{2}{3}
\fillcell{3}{1}
\fillcell{3}{2}
\fillcell{3}{3}
\setrow{3}{{\Graveltrap},{},{}}
\setrow{2}{{},{\Graveltrap},{}}
\setrow{1}{{},{},{\Graveltrap}}
\framepuzzle[black!60]
\end{logicpuzzle}
}
\subfloat[Logarithmic link, log rate ratio]{
\begin{logicpuzzle}[rows=3,columns=3,color=cyan!70, fontsize=small]
\valueH{MTE, CTEM, PACTE}
\valueV{PACTE,CTEM,MTE}
\fillcell{1}{3}
\fillcell{2}{1}
\fillcell{2}{2}
\fillcell{3}{1}
\fillcell{3}{2}
\setrow{3}{{\Graveltrap},{},{}}
\setrow{2}{{},{\Graveltrap},{}}
\setrow{1}{{},{},{\Graveltrap}}
\framepuzzle[black!60]
\end{logicpuzzle}
}
\subfloat[Logistic link, log odds ratio]{
\begin{logicpuzzle}[rows=3,columns=3,color=cyan!70, fontsize=small]
\valueH{MTE, CTEM, PACTE}
\valueV{PACTE,CTEM,MTE}
\fillcell{1}{3}
\fillcell{2}{1}
\fillcell{2}{2}
\fillcell{3}{1}
\fillcell{3}{2}
\setrow{3}{{\Graveltrap},{},{}}
\setrow{2}{{},{\Graveltrap},{}}
\setrow{1}{{},{},{\Graveltrap}}
\framepuzzle[black!60]
\end{logicpuzzle}
}
\caption{Matrices indicating whether different estimands are equivalent for the heterogeneous illustrative models. The blue squares denote matching estimand values; the dots denote the diagonal, where estimands are equivalent by definition.}
\label{matrices_heterogeneous}
\end{figure}

Table \ref{estimands_heterogeneous} presents a taxonomy of marginal estimands based on the heterogeneous illustrative models. Figure \ref{matrices_heterogeneous} indicates whether different estimands match for the heterogeneous illustrative models, across link functions and summary measures. The blue squares denote matching estimand values; the dots denote the diagonal, where logically the estimands are equivalent. In Table \ref{estimands_heterogeneous} and Figure \ref{matrices_heterogeneous}, the summary measures are on the additive linear predictor scale imposed by the link function used to model the corresponding outcome type.

\begin{table}[!htb]
\caption{Model-based marginal estimands for the heterogeneous illustrative models. For count outcomes and the log link, person-time is assumed constant, such that the log rate ratio is collapsible and can be interpreted as a log risk ratio.}
\centering
\scalebox{0.9}{
\begin{tabular}{llll}
\toprule
Outcome & Link function & Summary measure & 
Marginal estimand \\
\midrule
Continuous & Identity & Mean difference & Only depends on effect modifier means \\
Count & Logarithmic & Log risk ratio & \specialcell[t]{Depends on the full joint distribution of effect modifiers and \\ purely prognostic covariates that are associated with the former} \\
Binary & Logit & Log odds ratio & \specialcell[t]{Depends on the full joint distribution of effect modifiers and \\ purely prognostic covariates}\\
\bottomrule
\end{tabular}
}
\label{estimands_heterogeneous}
\end{table}

\subsection{Non-linear (quadratic) heterogeneous CTEX}\label{subsec33}

In the heterogeneous illustrative model in Section \ref{subsec32}, the conditional outcome expectation on the linear predictor scale varies linearly with $X$, in both the active treatment and the control groups. Hence, the CTEX -- the difference between the (transformed) conditional outcome expectations for each treatment -- is also a linear function of $X$. In practice, such outcome expectations may vary non-linearly with the covariate in at least one of the treatment groups, in which case the CTEX is a non-linear function of $X$. We now adopt another heterogeneous illustrative model, where the conditional outcome expectations on the linear predictor scale vary quadratically with $X$ in both treatment groups, such that the CTEX is a quadratic function of $X$. 

The \textit{quadratic} (heterogeneous) illustrative model specifies the conditional expectation of potential outcome $Y^t$ for treatment $T=t$ given $X$ as:
\begin{equation}
E (Y^{t} \mid X) = g^{-1} \left( \beta_0 + \beta_1 X + \beta_{2} X^2 + \beta_T t + \beta_{1T}X t + \beta_{2T}X^2 t \right). 
\label{eqn1_quadratic}
\end{equation}
Again, $X$ is prognostic of outcome and a (conditional) treatment effect modifier on the linear predictor scale. In this case, the effect modification mechanism is more complex \citep{mayer2017effect}. Due to the squared covariate-treatment product term, treatment efficacy depends on $X^2$. Substituting Equation \ref{eqn1_quadratic} into Equation \ref{cond_additive} yields $\textnormal{CTEX} = \beta_T + \beta_{1T} x + \beta_{2T} x^2 $ at $X=x$ on the linear predictor scale. That is, for the identity, logarithmic or logit link, the corresponding conditional mean difference, log risk ratio or log odds ratio, respectively, depends on the values of the covariate and the squared covariate. Applying Equation \ref{eqn1_quadratic} to Equation \ref{ctem} yields $\textnormal{CTEM}=\beta_T + \beta_{1T} \bar{X} + \beta_{2T} \bar{X}^2$, a quadratic function of the covariate mean. 

On the other hand, substituting Equation \ref{eqn1_quadratic} into Equation \ref{pacte} gives the population-average conditional estimand $\textnormal{PACTE}=E_X \left (\beta_T + \beta_{1T} X + \beta_{2T} X^2  \right ) = \beta_T +  \beta_{1T} E\left (  X \right )  + \beta_{2T} E \left ( X^2 \right ) = 
\beta_T +  \beta_{1T} \bar{X}  + \beta_{2T} E \left ( X^2 \right )$. The CTEM is no longer equal to the PACTE because the square of the covariate mean is not equal to the mean of the squared covariate. Because $f(X)=X^2$ is a convex function, due to Jensen's inequality: $E(X^2) \geq \bar{X}^2$. As both $E(X^2)$ and $\bar{X}^2$ are positive, the inequality implies that $\textnormal{PACTE} > \textnormal{CTEM}$ for $\beta_{2T} > 0$, and $\textnormal{PACTE} < \textnormal{CTEM}$ for $\beta_{2T} < 0$. 

With the quadratic illustrative model we want to show that, unlike for the model in Section \ref{subsec32}, the CTEM and the PACTE are in general not equivalent. When the CTEX is constant or a linear function of $X$, the PACTE expression happens to collapse to the CTEM on the linear predictor scale. Nevertheless, this is not the case when the CTEX is a non-linear function of $X$. Here, the interpretation of the CTEM is restricted to the covariate mean and no longer reflects a population-average conditional effect. Likewise, and contrary to suggestions in the literature \citep{remiro2021conflating}, the PACTE is a population-average estimand that does not necessarily represent the effect for subjects with the mean covariate values.

With respect to the marginal estimand, suppose $g(\cdot)$ is the identity link. On the mean difference scale, by the law of total expectation:
\begin{align*}
\textnormal{MTE} &= 
E \left(Y^1  \right)  - 
E \left(Y^0 \right) \\
&= E_X \left[ E \left( Y^1 \mid X \right) \right] -E_X \left [ E  \left(Y^0 \mid X \right) \right ] \\ &= 
E_X \left [\beta_0 + \beta_1 X + \beta_{2} X^2 \right ]
+ E_X \left [ \beta_T  + \beta_{1T} X + \beta_{2T} X^2  \right ] - E_X \left [\beta_0 + \beta_1 X + \beta_{2} X^2 \right] \\
&=\beta_T  + \beta_{1T} \bar{X} + \beta_{2T} E[X^2],
\end{align*}
such that the marginal treatment effect estimand coincides with the PACTE but not with the CTEM. Moreover, the formula $Var(X)=E(X^2) - \bar{X}^2$ implies that the marginal mean difference $\textnormal{MTE}=\beta_T  + \beta_{1T} \bar{X} + \beta_{2T} \left ( \bar{X}^2 + Var(X) \right )$. Consequently, and due to the direct collapsibility of the mean difference, the marginal estimand can be reduced to simple terms involving model coefficients and the mean and variance of the effect modifier $X$. 

Conversely, as per Section \ref{subsec32}, this is not the case for the marginal (log) risk ratio with $g(\cdot) = \log(\cdot)$, and for the marginal (log) odds ratio with $g(\cdot) = \logit(\cdot)$ -- nor for any other non-linear link function -- for which the MTE on the corresponding linear predictor scale is not equivalent to the CTEM or to the PACTE. Again, for these outcome-generating models and summary measures, the MTE depends on the full distribution of covariate $X$ (not just its mean and its variance), and also the distribution of baseline risk, as some prognostic factors may be unavailable. With multiple prognostic baseline covariates and in the presence of treatment effect heterogeneity, the MTE will generally depend on the full joint distribution of the effect modifiers and purely prognostic covariates \citep{remiro2024transportability}. 

\begin{figure}[!htb]
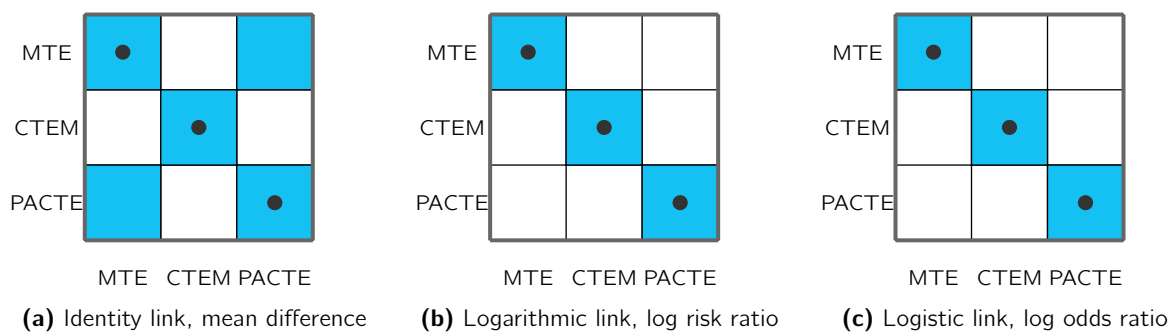

\centering
\subfloat[Identity link, mean difference]{
\begin{logicpuzzle}[rows=3,columns=3,color=cyan!70, fontsize=small]
\valueH{MTE, CTEM, PACTE}
\valueV{PACTE,CTEM,MTE}
\fillcell{1}{1}
\fillcell{1}{3}
\fillcell{2}{2}
\fillcell{3}{1}
\fillcell{3}{3}
\setrow{3}{{\Graveltrap},{},{}}
\setrow{2}{{},{\Graveltrap},{}}
\setrow{1}{{},{},{\Graveltrap}}
\framepuzzle[black!60]
\end{logicpuzzle}
}
\subfloat[Logarithmic link, log risk ratio]{
\begin{logicpuzzle}[rows=3,columns=3,color=cyan!70, fontsize=small]
\valueH{MTE, CTEM, PACTE}
\valueV{PACTE,CTEM,MTE}
\fillcell{1}{3}
\fillcell{2}{2}
\fillcell{3}{1}
\setrow{3}{{\Graveltrap},{},{}}
\setrow{2}{{},{\Graveltrap},{}}
\setrow{1}{{},{},{\Graveltrap}}
\framepuzzle[black!60]
\end{logicpuzzle}
}
\subfloat[Logistic link, log odds ratio]{
\begin{logicpuzzle}[rows=3,columns=3,color=cyan!70, fontsize=small]
\valueH{MTE, CTEM, PACTE}
\valueV{PACTE,CTEM,MTE}
\fillcell{1}{3}
\fillcell{2}{2}
\fillcell{3}{1}
\setrow{3}{{\Graveltrap},{},{}}
\setrow{2}{{},{\Graveltrap},{}}
\setrow{1}{{},{},{\Graveltrap}}
\framepuzzle[black!60]
\end{logicpuzzle}
}
\caption{Matrices indicating whether different estimands are equivalent for the quadratic (heterogeneous) illustrative models. The blue squares denote matching estimand values; the dots denote the diagonal, where estimands are equivalent by definition.}
\label{matrices_quadratic}
\end{figure}

Table \ref{estimands_quadratic} presents a taxonomy of marginal estimands based on the quadratic illustrative models. Figure \ref{matrices_quadratic} indicates whether different estimands match for the quadratic illustrative models, across link functions and summary measures. The blue squares denote matching estimand values; the dots denote the diagonal, where logically the estimands are equivalent. In Table \ref{estimands_quadratic} and Figure \ref{matrices_quadratic}, the summary measures are on the additive linear predictor scale imposed by the link function used to model the corresponding outcome type.

\begin{table}[!htb]
\caption{Model-based marginal estimands for the quadratic (heterogeneous) illustrative models. For count outcomes and the log link, person-time is assumed constant, such that the log rate ratio is collapsible and can be interpreted as a log risk ratio.}
\centering
\scalebox{0.9}{
\begin{tabular}{llll}
\toprule
Outcome & Link function & Summary measure & 
Marginal estimand \\
\midrule
Continuous & Identity & Mean difference & Depends on effect modifier means and variances \\
Count & Logarithmic & Log risk ratio & \specialcell[t]{Depends on the full joint distribution of effect modifiers and \\ purely prognostic covariates that are associated with the former} \\
Binary & Logit & Log odds ratio & \specialcell[t]{Depends on the full joint distribution of effect modifiers and \\ purely prognostic covariates}\\
\bottomrule
\end{tabular}
}
\label{estimands_quadratic}
\end{table}

\section{Indirect treatment comparisons}

We have clarified the interpretation of different marginal and conditional summary measures, considering different types of outcomes and outcome-generating mechanisms, with and without effect modification. We illustrate potential implications of estimand incompatibility for evidence synthesis on the following two-study scenario, increasingly encountered in health technology assessment \citep{phillippo2016nice, phillippo2018methods}. 

A manufacturer submitting evidence for reimbursement has conducted its own ``index'' RCT comparing the efficacy of novel active treatment $A$ versus control $C$. Another company developed earlier a competing active treatment and performed an RCT (the ``competitor'' trial) comparing the efficacy of its product, active treatment $B$, against control $C$. An \textit{indirect treatment comparison} between $A$ and $B$, \textit{anchored} at the common comparator $C$, is required for reimbursement purposes. Typically, the manufacturer submitting the evidence has individual patient data (IPD) from the index RCT, but only aggregate-level summary data from the competitor RCT. For the latter trial, access to subject-level data is unavailable due to privacy and confidentiality concerns, which restricts the target population of the analysis to the competitor trial population.  

Covariate adjustment methods -- denoted ``population adjustment'' in this context -- have been proposed to project the treatment effect for $A$ versus $C$ from the index trial to the competitor trial. A ``fair'' indirect comparison is then performed in the competitor trial population. Namely, the treatment effect for $A$ versus $B$ is estimated by combining treatment effect estimates for $A$ versus $C$ and for $B$ versus $C$, on the additive scale: 
\begin{equation}
\hat{\Delta}_{AB}^{(BC)} = \hat{\Delta}_{AC}^{(BC)} - \hat{\Delta}_{BC}^{(BC)},
\label{anchored_ITC}
\end{equation}
where $\hat{\Delta}_{tt'}^{(BC)}$ is the effect estimate for $t$ versus $t'$, with superscript $BC$ indexing the competitor trial population. 

To produce the estimate $\hat{\Delta}_{AC}^{(BC)}$ in Equation \ref{anchored_ITC}, several population adjustment methods have been developed: matching-adjusted indirect comparison (MAIC) \citep{signorovitch2010comparative}, simulated treatment comparison (STC), multilevel network meta-regression (ML-NMR) \citep{phillippo2020multilevel}, network meta-interpolation (NMI) \citep{harari2023network} and cross-network meta-regression (cross-NMR) \citep{hamza2023synthesizing, hamza2024crossnma}. An oft-overlooked aspect is that these methods estimate different summary measures, as outlined in Table \ref{methodology_estimands}. The ``plug-in'' or ``mean-centering'' version of STC (STC-P) \citep{phillippo2016nice}, NMI and cross-NMR estimate the CTEM; MAIC and simulation- or G-computation-based adaptations of STC (STC-G) \citep{remiro2022parametric} estimate the MTE; and ML-NMR can estimate the PACTE or the MTE \citep{phillippo2023validating}.

\begin{table}[!htb]
\caption{Summary measures estimated by different covariate adjustment approaches in the context of indirect comparisons.}
\centering
\begin{tabular}{lll}
\toprule
Methodology & Summary measure \\
\midrule
Matching-adjusted indirect comparison (MAIC) & MTE  \\
``Plug-in'' simulated treatment comparison (STC-P) & CTEM \\
``G-computation'' simulated treatment comparison (STC-G) & MTE \\
Multilevel network meta-regression (ML-NMR)& PACTE and MTE \\
Network meta-interpolation (NMI) & CTEM \\
Cross-network meta-regression (cross-NMR) & CTEM \\
\bottomrule
\end{tabular}
\label{methodology_estimands}
\end{table}

In MAIC and STC-G, $\hat{\Delta}_{AC}^{(BC)}$ is an estimate of the MTE. As such, the $\hat{\Delta}_{BC}^{(BC)}$ plugged into Equation \ref{anchored_ITC} must estimate a compatible marginal summary measure. This is not problematic; a point estimate of the MTE is typically reported in RCT publications. Alternatively, ``crude'' unadjusted point estimates of the MTE can be derived from published aggregate-level outcome data, e.g., for binary outcomes, marginal risk differences, log risk ratios or log odds ratios can be calculated from event counts reported in $2\times2$ contingency tables. 

STC-P, NMI and cross-NMR substitute mean covariate values for the competitor trial population when estimating the covariate-adjusted treatment effect for $A$ versus $C$, such that $\hat{\Delta}_{AC}^{(BC)}$ is an estimate of the CTEM. However, any estimate $\hat{\Delta}_{BC}^{(BC)}$ that is available for $B$ versus $C$ is likely an incompatible marginal or conditional summary measure. RCTs often report estimates of conditional estimands, e.g., where the published effect estimate is the treatment coefficient of a multivariable regression of outcome on treatment and baseline covariates. For simplicity, our exposition in Section \ref{sec2} and Section \ref{sec3} considered a single baseline covariate, and all conditional summary measures condition on such covariate. However, conditioning on different sets of covariates leads to different conditional summary measures or estimands. 

Trials with conditional estimates are likely to report effects conditioned on different covariate sets. These may be incompatible, especially for non-collapsible measures, even if the trials have identical target populations \citep{daniel2021making}. For a given target population, there are a multitude of PACTEs and CTEMs, one for every combination of baseline covariates used for conditioning. For STC-P, NMI and cross-NMR, if conditional estimates are reported for the competitor trial, $\hat{\Delta}_{AC}^{(BC)}$ and $\hat{\Delta}_{BC}^{(BC)}$ would have to condition on the same covariate set to guarantee estimand compatibility. 

ML-NMR is not susceptible to estimand incompatibility issues. The multilevel regression allows evidence to be combined at the level of the CTEX, directly producing CTEX estimates for $A$ versus $B$. PACTE estimates for $A$ versus $B$ are obtained by averaging the CTEX on the linear predictor scale. The individual-level outcome model can be used to predict outcomes under treatments $A$ and $B$, on their natural scale. Averaging the predictions yields marginal outcome mean estimates under each treatment, and one can produce an MTE estimate for $A$ versus $B$ by contrasting the back-transformed marginal outcome means on the scale of interest \citep{phillippo2023validating}. 

Population-level decision-making requires population-average estimates that are relevant to the decision target population. The target population need not be represented by that of the competitor trial or may not be reflected by any of the studies in the evidence base. An advantage of ML-NMR is that it is not limited to produce treatment effect estimates in the competitor trial population, and can produce population-average marginal and conditional estimates in any target population of interest. This typically requires an additional identifying assumption, which may be relaxed or assessed in larger evidence networks.\citep{phillippo2020multilevel, phillippo2023validating}.

\section{Discussion and concluding remarks}

While ICH E9 (R1) lacks explicit guidelines for estimands in evidence synthesis, it warns of potential estimand incompatibilities, cautioning against na\"ive comparisons across trials without considering the estimand that is addressed in each. 

We have clarified the interpretation of different marginal and conditional summary measures, and illustrated practical implications of estimand incompatibility for indirect treatment comparisons. Several novel covariate adjustment methodologies have been developed in this area, with different approaches targeting different summary measures. The estimand incompatibility issues described in this article may also apply to other areas of evidence synthesis, such as the na\"ive pooling of marginal and conditional estimates in pairwise -- unadjusted or covariate-adjusted -- meta-analyses of direct treatment comparisons \citep{hedges2016network}, and across indirect treatment comparisons or network meta-analyses that do not adjust for covariates \citep{rucker2014network}. 

Estimand incompatibility is particularly concerning for evidence synthesis based on published summary effect measures, as different trials may report different summary measures. Full access to subject-level data for all trials is rarely available in practice. However, this would be beneficial to resolve estimand incompatibility issues by allowing for an IPD (network) meta-regression, the ``gold standard'' evidence synthesis approach that ML-NMR seeks to generalize. Such an approach can produce any desired marginal or conditional summary measure without incompatibility issues, resolve heterogeneity due to population differences, and perform treatment comparisons in any given target population of interest.

Many evidence syntheses of aggregate-level data are based on raw outcome data reported in each study arm, e.g., event counts and sample sizes, and thus combine marginal treatment effects across studies. However, purely prognostic covariates that are not conventionally described as effect modifiers – which do not interact with treatment nor induce treatment effect heterogeneity across subjects – can modify marginal treatment effects and compromise transportability, particularly for non-collapsible summary measures. This is generally under-appreciated in the evidence synthesis literature, with the exception of a recent article on network meta-analysis \citep{riley2023using}. Whether particular covariates can be classed as effect modifiers -- thereby leading to departures from transportability, constancy or external validity -- depends on the summary measure, and on whether it is marginal or conditional.  

While the summary measure is at the heart of the estimand framework, as described by ICH E9 (R1), it is not an explicit component of frameworks such as PICO (Population, Intervention, Comparator, Outcome), typically used for scoping in evidence synthesis, and often a secondary consideration when postulating research questions in evidence synthesis. We emphasize the importance of the summary measure as a fundamental component of any clinical research question.

Within evidence synthesis, we can think of estimands at two different hierarchical levels: at the trial level and at the meta-analytical level \citep{remiro2022some}. Both levels are important. There is a need to align estimands across different studies, as highlighted by this article. There is also a need to align the estimand targeted at the meta-analytical level with a relevant clinical research question, motivating the shift towards more targeted meta-analyses that enhance applicability with respect to healthcare decision-making contexts \citep{ades2024twenty, remiro2024broad}. 

\ack{Not available.}

\section*{Related articles}

\cite{hedges2016network}, \cite{greenland2014effect}, \cite{greenland2015collapsibility}, \cite{mcknight2015effect},
\cite{rucker2014network}.

\appendix

%% In using BibTeX, use wb_stat.bst
\bibliographystyle{wb_stat}
\bibliography{WileySTAT}

\begin{thebibliography}{48}
\newcommand{\enquote}[1]{`#1'}
\providecommand{\natexlab}[1]{#1}
\expandafter\ifx\csname urlstyle\endcsname\relax
  \providecommand{\doi}[1]{doi:\discretionary{}{}{}#1}\else
  \providecommand{\doi}{doi:\discretionary{}{}{}\begingroup \urlstyle{rm}\Url}\fi

\bibitem[{Ades et~al.(2024)Ades, Welton, Dias, Phillippo \& Caldwell}]{ades2024twenty}
Ades, A, Welton, NJ, Dias, S, Phillippo, DM \& Caldwell, DM (2024), \enquote{Twenty years of network meta-analysis: Continuing controversies and recent developments,} \emph{Research Synthesis Methods}, \textbf{15}(5), pp. 702--727.

\bibitem[{Arel-Bundock et~al.(2024)Arel-Bundock, Greifer \& Heiss}]{arel2024interpret}
Arel-Bundock, V, Greifer, N \& Heiss, A (2024), \enquote{How to interpret statistical models using marginaleffects for {R} and {P}ython,} \emph{Journal of Statistical Software}, \textbf{111}, pp. 1--32.

\bibitem[{Austin(2013)}]{austin2013performance}
Austin, PC (2013), \enquote{The performance of different propensity score methods for estimating marginal hazard ratios,} \emph{Statistics in Medicine}, \textbf{32}(16), pp. 2837--2849.

\bibitem[{Austin(2014)}]{austin2014use}
Austin, PC (2014), \enquote{The use of propensity score methods with survival or time-to-event outcomes: reporting measures of effect similar to those used in randomized experiments,} \emph{Statistics in Medicine}, \textbf{33}(7), pp. 1242--1258.

\bibitem[{Ballman(2015)}]{ballman2015biomarker}
Ballman, KV (2015), \enquote{Biomarker: predictive or prognostic?} \emph{Journal of clinical oncology: official journal of the American Society of Clinical Oncology}, \textbf{33}(33), pp. 3968--3971.

\bibitem[{Christensen et~al.(2021)Christensen, Bours \& Nielsen}]{christensen2021effect}
Christensen, R, Bours, MJ \& Nielsen, SM (2021), \enquote{Effect modifiers and statistical tests for interaction in randomized trials,} \emph{Journal of Clinical Epidemiology}, \textbf{134}, pp. 174--177.

\bibitem[{Colnet et~al.(2023)Colnet, Josse, Varoquaux \& Scornet}]{colnet2023risk}
Colnet, B, Josse, J, Varoquaux, G \& Scornet, E (2023), \enquote{Risk ratio, odds ratio, risk difference... which causal measure is easier to generalize?} \emph{arXiv preprint arXiv:2303.16008}.

\bibitem[{Daniel et~al.(2021)Daniel, Zhang \& Farewell}]{daniel2021making}
Daniel, R, Zhang, J \& Farewell, D (2021), \enquote{Making apples from oranges: Comparing noncollapsible effect estimators and their standard errors after adjustment for different covariate sets,} \emph{Biometrical Journal}, \textbf{63}(3), pp. 528--557.

\bibitem[{Hamza et~al.(2023)Hamza, Chalkou, Pellegrini, Kuhle, Benkert, Lorscheider, Zecca, Iglesias-Urrutia, Manca, Furukawa et~al.}]{hamza2023synthesizing}
Hamza, T, Chalkou, K, Pellegrini, F, Kuhle, J, Benkert, P, Lorscheider, J, Zecca, C, Iglesias-Urrutia, CP, Manca, A, Furukawa, TA et~al. (2023), \enquote{Synthesizing cross-design evidence and cross-format data using network meta-regression,} \emph{Research Synthesis Methods}, \textbf{14}(2), pp. 283--300.

\bibitem[{Hamza et~al.(2024)Hamza, Schwarzer \& Salanti}]{hamza2024crossnma}
Hamza, T, Schwarzer, G \& Salanti, G (2024), \enquote{crossnma: An {R} package to synthesize cross-design evidence and cross-format data using network meta-analysis and network meta-regression,} \emph{BMC Medical Research Methodology}, \textbf{24}(1), pp. 1--16.

\bibitem[{Harari et~al.(2023)Harari, Soltanifar, Cappelleri, Verhoek, Ouwens, Daly \& Heeg}]{harari2023network}
Harari, O, Soltanifar, M, Cappelleri, JC, Verhoek, A, Ouwens, M, Daly, C \& Heeg, B (2023), \enquote{Network meta-interpolation: Effect modification adjustment in network meta-analysis using subgroup analyses,} \emph{Research Synthesis Methods}, \textbf{14}(2), pp. 211--233.

\bibitem[{Hern{\'a}n \& Robins(2020)}]{hernan2020causal}
Hern{\'a}n, MA \& Robins, JM (2020), \emph{Causal inference: what if}, Boca Raton: Chapman \& Hall/CRC.

\bibitem[{H{\o}jbjerre-Frandsen et~al.(2025)H{\o}jbjerre-Frandsen, van~der Laan \& Schuler}]{hojbjerre2025powering}
H{\o}jbjerre-Frandsen, E, van~der Laan, MJ \& Schuler, A (2025), \enquote{Powering {RCT}s for marginal effects with {GLM}s using prognostic score adjustment,} \emph{arXiv preprint arXiv:2503.22284}.

\bibitem[{Huitfeldt et~al.(2019)Huitfeldt, Stensrud \& Suzuki}]{huitfeldt2019collapsibility}
Huitfeldt, A, Stensrud, MJ \& Suzuki, E (2019), \enquote{On the collapsibility of measures of effect in the counterfactual causal framework,} \emph{Emerging Themes in Epidemiology}, \textbf{16}(1), pp. 1--5.

\bibitem[{Keene et~al.(2023)Keene, Lynggaard, Englert, Lanius \& Wright}]{keene2023estimands}
Keene, ON, Lynggaard, H, Englert, S, Lanius, V \& Wright, D (2023), \enquote{Why estimands are needed to define treatment effects in clinical trials,} \emph{BMC Medicine}, \textbf{21}(1), p. 276.

\bibitem[{Kiefer \& Mayer(2019)}]{kiefer2019average}
Kiefer, C \& Mayer, A (2019), \enquote{Average effects based on regressions with a logarithmic link function: A new approach with stochastic covariates,} \emph{Psychometrika}, \textbf{84}(2), pp. 422--446.

\bibitem[{Lenth(2016)}]{lenth2016least}
Lenth, RV (2016), \enquote{Least-squares means: the {R} package lsmeans,} \emph{Journal of Statistical Software}, \textbf{69}(1), pp. 1--33.

\bibitem[{Martinussen \& Vansteelandt(2013)}]{martinussen2013collapsibility}
Martinussen, T \& Vansteelandt, S (2013), \enquote{On collapsibility and confounding bias in {C}ox and {A}alen regression models,} \emph{Lifetime Data Analysis}, \textbf{19}(3), pp. 279--296.

\bibitem[{Mayer et~al.(2017)Mayer, Umbach, Flunger \& Kelava}]{mayer2017effect}
Mayer, A, Umbach, N, Flunger, B \& Kelava, A (2017), \enquote{Effect analysis using nonlinear structural equation mixture modeling,} \emph{Structural Equation Modeling: A Multidisciplinary Journal}, \textbf{24}(4), pp. 556--570.

\bibitem[{Morris et~al.(2022)Morris, Walker, Williamson \& White}]{morris2022planning}
Morris, TP, Walker, AS, Williamson, EJ \& White, IR (2022), \enquote{Planning a method for covariate adjustment in individually randomised trials: a practical guide,} \emph{Trials}, \textbf{23}(1), pp. 1--17.

\bibitem[{M{\"u}tze et~al.(2025)M{\"u}tze, Bell, Englert, Hougaard, Jackson, Lanius \& Ravn}]{mutze2025principles}
M{\"u}tze, T, Bell, J, Englert, S, Hougaard, P, Jackson, D, Lanius, V \& Ravn, H (2025), \enquote{Principles for defining estimands in clinical trials—a proposal,} \emph{Pharmaceutical Statistics}, \textbf{24}(1), p. e2432.

\bibitem[{Phillippo et~al.(2016)Phillippo, Ades, Dias, Palmer, Abrams \& Welton}]{phillippo2016nice}
Phillippo, D, Ades, T, Dias, S, Palmer, S, Abrams, KR \& Welton, N (2016), \enquote{{NICE} {DSU} technical support document 18: methods for population-adjusted indirect comparisons in submissions to {NICE},} \emph{NICE Decision Support Unit}.

\bibitem[{Phillippo et~al.(2018)Phillippo, Ades, Dias, Palmer, Abrams \& Welton}]{phillippo2018methods}
Phillippo, DM, Ades, AE, Dias, S, Palmer, S, Abrams, KR \& Welton, NJ (2018), \enquote{Methods for population-adjusted indirect comparisons in health technology appraisal,} \emph{Medical Decision Making}, \textbf{38}(2), pp. 200--211.

\bibitem[{Phillippo et~al.(2023)Phillippo, Dias, Ades, Belger, Brnabic, Saure, Schymura \& Welton}]{phillippo2023validating}
Phillippo, DM, Dias, S, Ades, A, Belger, M, Brnabic, A, Saure, D, Schymura, Y \& Welton, NJ (2023), \enquote{Validating the assumptions of population adjustment: application of multilevel network meta-regression to a network of treatments for plaque psoriasis,} \emph{Medical Decision Making}, \textbf{43}(1), pp. 53--67.

\bibitem[{Phillippo et~al.(2020)Phillippo, Dias, Ades, Belger, Brnabic, Schacht, Saure, Kadziola \& Welton}]{phillippo2020multilevel}
Phillippo, DM, Dias, S, Ades, A, Belger, M, Brnabic, A, Schacht, A, Saure, D, Kadziola, Z \& Welton, NJ (2020), \enquote{Multilevel network meta-regression for population-adjusted treatment comparisons,} \emph{Journal of the Royal Statistical Society: Series A (Statistics in Society)}, \textbf{183}(3), pp. 1189--1210.

\bibitem[{Phillippo et~al.(2021)Phillippo, Dias, Ades \& Welton}]{phillippo2021target}
Phillippo, DM, Dias, S, Ades, AE \& Welton, NJ (2021), \enquote{Target estimands for efficient decision making: Response to comments on “{A}ssessing the performance of population adjustment methods for anchored indirect comparisons: A simulation study”,} \emph{Statistics in Medicine}, \textbf{40}(11), pp. 2759--2763.

\bibitem[{Phillippo et~al.(2025)Phillippo, Remiro-Az{\'o}car, Heath, Baio, Dias, Ades \& Welton}]{Phillippo_Remiro-Azócar_Heath_Baio_Dias_Ades_Welton_2025}
Phillippo, DM, Remiro-Az{\'o}car, A, Heath, A, Baio, G, Dias, S, Ades, A \& Welton, NJ (2025), \enquote{Effect modification and non-collapsibility together may lead to conflicting treatment decisions: A review of marginal and conditional estimands and recommendations for decision-making,} \emph{Research Synthesis Methods}, \textbf{16}(2), pp. 1--27.

\bibitem[{Remiro-Az{\'o}car(2022{\natexlab{a}})}]{remiro2022some}
Remiro-Az{\'o}car, A (2022{\natexlab{a}}), \enquote{Some considerations on target estimands for health technology assessment,} \emph{Statistics in Medicine}, \textbf{41}(28), pp. 5592--5596.

\bibitem[{Remiro-Az{\'o}car(2022{\natexlab{b}})}]{remiro2022target}
Remiro-Az{\'o}car, A (2022{\natexlab{b}}), \enquote{Target estimands for population-adjusted indirect comparisons,} \emph{Statistics in Medicine}, \textbf{41}(28), pp. 5558--5569.

\bibitem[{Remiro-Az{\'o}car(2024)}]{remiro2024transportability}
Remiro-Az{\'o}car, A (2024), \enquote{Transportability of model-based estimands in evidence synthesis,} \emph{Statistics in Medicine}, \textbf{43}(22), pp. 4217--4249.

\bibitem[{Remiro-Az{\'o}car \& Gorst-Rasmussen(2024)}]{remiro2024broad}
Remiro-Az{\'o}car, A \& Gorst-Rasmussen, A (2024), \enquote{Broad versus narrow research questions in evidence synthesis: a parallel to (and plea for) estimands,} \emph{Research Synthesis Methods}, \textbf{15}(5), pp. 735--740.

\bibitem[{Remiro-Az{\'o}car et~al.(2021{\natexlab{a}})Remiro-Az{\'o}car, Heath \& Baio}]{remiro2021conflating}
Remiro-Az{\'o}car, A, Heath, A \& Baio, G (2021{\natexlab{a}}), \enquote{Conflating marginal and conditional treatment effects: Comments on “{A}ssessing the performance of population adjustment methods for anchored indirect comparisons: A simulation study”,} \emph{Statistics in Medicine}, \textbf{40}(11), pp. 2753--2758.

\bibitem[{Remiro-Az{\'o}car et~al.(2021{\natexlab{b}})Remiro-Az{\'o}car, Heath \& Baio}]{remiro2021methods}
Remiro-Az{\'o}car, A, Heath, A \& Baio, G (2021{\natexlab{b}}), \enquote{Methods for population adjustment with limited access to individual patient data: A review and simulation study,} \emph{Research Synthesis Methods}, \textbf{12}(6), pp. 750--775.

\bibitem[{Remiro-Az{\'o}car et~al.(2022)Remiro-Az{\'o}car, Heath \& Baio}]{remiro2022parametric}
Remiro-Az{\'o}car, A, Heath, A \& Baio, G (2022), \enquote{Parametric g-computation for compatible indirect treatment comparisons with limited individual patient data,} \emph{Research Synthesis Methods}, \textbf{13}(6), pp. 716--744.

\bibitem[{Riley et~al.(2023)Riley, Dias, Donegan, Tierney, Stewart, Efthimiou \& Phillippo}]{riley2023using}
Riley, RD, Dias, S, Donegan, S, Tierney, JF, Stewart, LA, Efthimiou, O \& Phillippo, DM (2023), \enquote{Using individual participant data to improve network meta-analysis projects,} \emph{BMJ Evidence-Based Medicine}, \textbf{28}(3), pp. 197--203.

\bibitem[{Signorovitch et~al.(2010)Signorovitch, Wu, Andrew, Gerrits, Kantor, Bao, Gupta \& Mulani}]{signorovitch2010comparative}
Signorovitch, JE, Wu, EQ, Andrew, PY, Gerrits, CM, Kantor, E, Bao, Y, Gupta, SR \& Mulani, PM (2010), \enquote{Comparative effectiveness without head-to-head trials,} \emph{Pharmacoeconomics}, \textbf{28}(10), pp. 935--945.

\bibitem[{Sj{\"o}lander et~al.(2016)Sj{\"o}lander, Dahlqwist \& Zetterqvist}]{sjolander2016note}
Sj{\"o}lander, A, Dahlqwist, E \& Zetterqvist, J (2016), \enquote{A note on the noncollapsibility of rate differences and rate ratios,} \emph{Epidemiology}, \textbf{27}(3), pp. 356--359.

\bibitem[{stat00513(2016)}]{hedges2016network}
stat00513 (2016), \enquote{Meta-analysis,} \emph{Hedges, L. Wiley StatsRef: Statistics Reference Online}.

\bibitem[{stat03728(2014)}]{greenland2014effect}
stat03728 (2014), \enquote{Effect modification and interaction,} \emph{Greenland, S. Wiley StatsRef: Statistics Reference Online}.

\bibitem[{stat05130(2015)}]{greenland2015collapsibility}
stat05130 (2015), \enquote{Collapsibility,} \emph{Greenland, S. Wiley StatsRef: Statistics Reference Online}.

\bibitem[{stat05152(2015)}]{mcknight2015effect}
stat05152 (2015), \enquote{Effect modification,} \emph{McKnight, B. Wiley StatsRef: Statistics Reference Online}.

\bibitem[{stat07909(2014)}]{rucker2014network}
stat07909 (2014), \enquote{Network meta-analysis,} \emph{R\"ucker, G. Wiley StatsRef: Statistics Reference Online}.

\bibitem[{Van~Lancker et~al.(2024)Van~Lancker, Bretz \& Dukes}]{van2023use}
Van~Lancker, K, Bretz, F \& Dukes, O (2024), \enquote{Covariate adjustment in randomized controlled trials: General concepts and practical considerations,} \emph{Clinical Trials}, \textbf{21}(4), pp. 399--411.

\bibitem[{Van~Lancker et~al.(2022)Van~Lancker, Vo \& Akacha}]{van2022estimands}
Van~Lancker, K, Vo, TT \& Akacha, M (2022), \enquote{Estimands in heath technology assessment: a causal inference perspective,} \emph{Statistics in Medicine}, \textbf{41}(28), pp. 5577--5585.

\bibitem[{Vansteelandt \& Dukes(2022)}]{vansteelandt2022assumption}
Vansteelandt, S \& Dukes, O (2022), \enquote{Assumption-lean inference for generalised linear model parameters,} \emph{Journal of the Royal Statistical Society Series B: Statistical Methodology}, \textbf{84}(3), pp. 657--685.

\bibitem[{Vellaisamy \& Vijay(2008)}]{vellaisamy2008collapsibility}
Vellaisamy, P \& Vijay, V (2008), \enquote{Collapsibility of regression coefficients and its extensions,} \emph{Journal of Statistical Planning and Inference}, \textbf{138}(4), pp. 982--994.

\bibitem[{Wei et~al.(2024)Wei, Xu, Bornkamp, Lin, Tian, Xi, Zhang, Zhao \& Roychoudhury}]{wei2024conditional}
Wei, J, Xu, J, Bornkamp, B, Lin, R, Tian, H, Xi, D, Zhang, X, Zhao, Z \& Roychoudhury, S (2024), \enquote{Conditional and unconditional treatment effects in randomized clinical trials: Estimands, estimation, and interpretation,} \emph{Statistics in Biopharmaceutical Research}, \textbf{16}(3), pp. 371--381.

\bibitem[{Westreich et~al.(2019)Westreich, Edwards, Lesko, Cole \& Stuart}]{westreich2019target}
Westreich, D, Edwards, JK, Lesko, CR, Cole, SR \& Stuart, EA (2019), \enquote{Target validity and the hierarchy of study designs,} \emph{American Journal of Epidemiology}, \textbf{188}(2), pp. 438--443.

\end{thebibliography}

\end{document}